\documentclass{article}

\usepackage{arxiv}

\usepackage[utf8]{inputenc} 
\usepackage[T1]{fontenc}    
\usepackage{hyperref}       
\usepackage{url}            
\usepackage{booktabs}       
\usepackage{amsfonts}       
\usepackage{nicefrac}       
\usepackage{microtype}      
\usepackage{graphicx}
\graphicspath{ {./images/} }
\usepackage{cite}
\usepackage{subfig}
\usepackage{xcolor}
\usepackage[flushleft]{threeparttable}

\title{Is disruption decreasing, or is it accelerating?}

\author{
 R. Alexander Bentley \\
College of Emerging and Collaborative Studies \\
University of Tennessee \\
Knoxville, TN 37996, USA. 
\texttt{rabentley@utk.edu} \\
   \And
 Sergi Valverde \\
 Institute of Evolutionary Biology \\
 Spanish National Research Council \\
 Barcelona, 08003, Spain 
 \And
Joshua Borycz \\ 
Stevenson Science and Engineering Library \\
Vanderbilt University \\
Nashville, TN 37203, USA 
 \And
Blai Vidiella \\
Institute of Evolutionary Biology \\
Spanish National Research Council \\
Barcelona, 08003, Spain
 \And
Benjamin D. Horne \\
School of Information Sciences \\
University of Tennessee \\
Knoxville, TN, 37996, USA
 \And
Salva Duran-Nebreda \\
Institute of Evolutionary Biology \\
Spanish National Research Council \\
Barcelona, 08003, Spain
 \And
Michael J. O'Brien \\
Dept. of Communication, History, Philosophy \\
Dept. of Life Sciences \\
Texas A\&M University \\
San Antonio, TX, 78224, USA
}

\begin{document}
\maketitle
\begin{abstract}
A recent highly-publicized study \cite{Park_etal_2023}, claiming that science has become less disruptive over recent decades, represents an extraordinary achievement but with deceptive results. The measure of disruption, CD$_5$, in this study \cite{Park_etal_2023} does not account for differences in citation amid decades of exponential growth in publication rate. In order to account for both the exponential growth as well as the differential impact of research works over time, here we apply a weighted disruption index to the same dataset. We find that, among research papers in the dataset \cite{Park_etal_2023}, this weighted disruption index has been close to its expected neutral value over the last fifty years and has even increased modestly since 2000. We also show how the proportional decrease in unique words \cite{Park_etal_2023} is expected in an exponentially growing corpus. Finding little evidence for recent decrease in disruption, we suggest that it is actually increasing. Future research should investigate improved definitions of disruption.
.
\end{abstract}


\section{Introduction}
Scientific research has become less disruptive according to a study \cite{Park_etal_2023} of the citation patterns among millions of published papers as well as patents. This result is based on the observed decrease in both a metric for disruption and the relative fraction of unique words over the last seventy years \cite{Park_etal_2023}. 

Scientific metrics provide evidence about a certain element of reality, but may not always reflect trends accurately. The reported decline in the disruption index, CD$_5$, since 1950 \cite{Park_etal_2023} is largely due to the diluting effect of exponential growth in quantity of scientific papers and patent applications, as well as their respective citations, over those decades. 

\begin{figure}[h!]
\centering
\includegraphics[width=.6\textwidth]{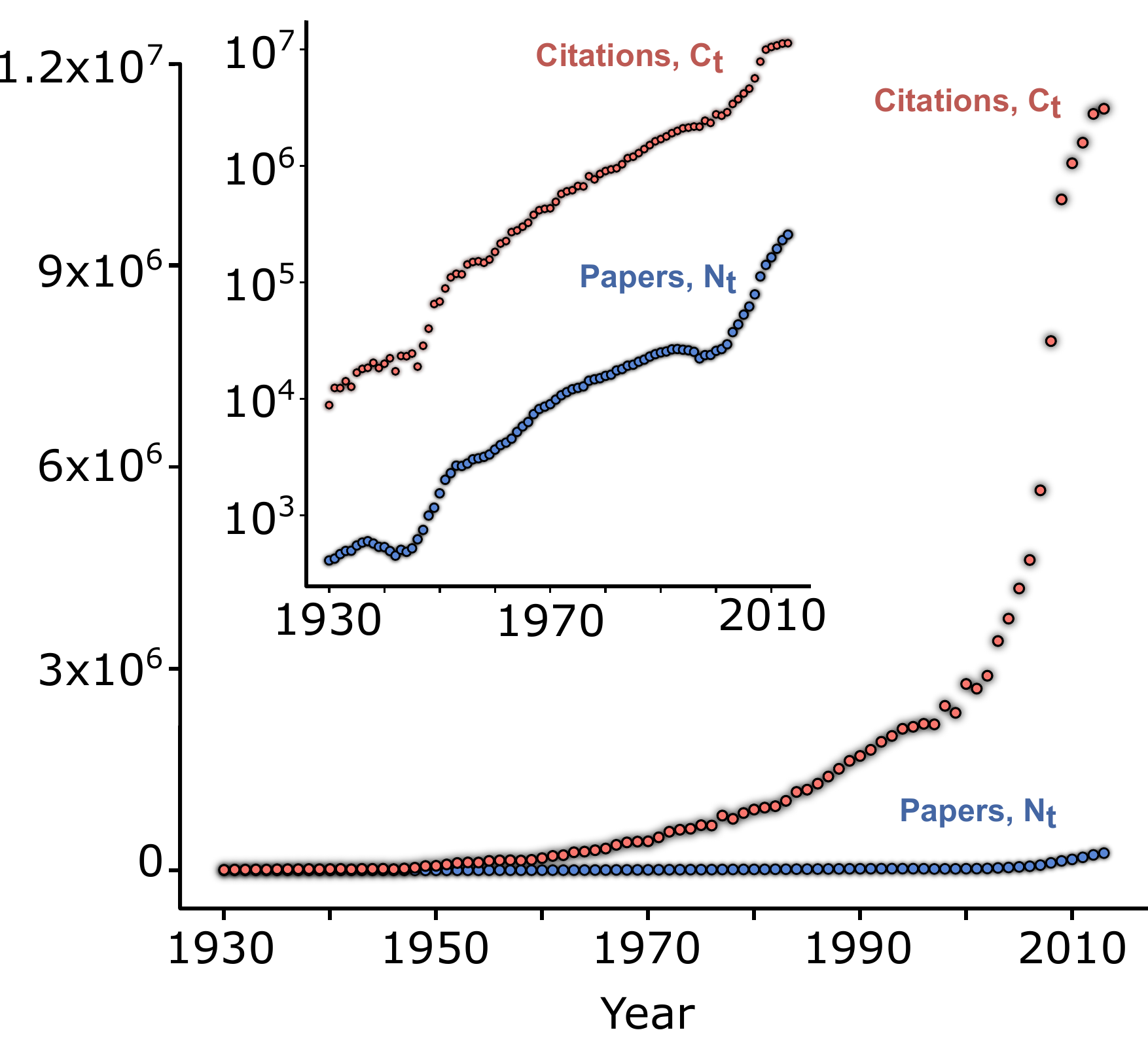}  
\caption{\textbf{Decades of exponential growth in scientific publications and citations:} $C_t$ and $N_t$ are annual numbers of citations and publications, respectively. Inset shows logarithmic scale. Source data from \cite{Park_etal_2023}.}
\label{fig1}
\end{figure}

In these results \cite{Park_etal_2023}, CD$_5$ measures the proportion of citations that a focal work receives relative to the works cited by the focal work. This index does not account for exponential growth of publications and citations. It also treats all papers equally in the annual average, such that a research paper cited by, say, just two other papers with a CD$_5$ score of $-1$ would offset a paper from the same year cited 1,000 times with a CD$_5$ score of $+1$. Averaged across an exponentially-increasing number of papers each year, this annually-averaged CD$_5$ has decreased in concave-up fashion \cite{Park_etal_2023}. This does not necessarily signal a decrease in disruption as defined by how much subsequent work is affected at an accelerating pace.

This highlights the need to account for decades of growth in numbers of papers (and patents) as well as a corpus of expanding citations. For example, the number of patents in the United States has grown exponentially since the 19th century (\url{www.uspto.gov/patents/milestones}). In academic publications, this growth rate averaged about $4\%$ per year for more than a century and about $5\%$ per year since 1950 \cite{Bornmann_etal_2020}. Accelerated growth appears to be responsible for the steady decline in disruption as defined in the study \cite{Park_etal_2023} because the number of citations is increasing in a similar way (Fig.~\ref{fig1}).  We confirm that the annual numbers of research papers in the data-set increased exponentially for over eight decades since 1930 (Fig.~\ref{fig1}, blue).  

\section{A citation-weighted measure of disruption}

 Using the term `paper' for simplicity, the disruption metric used in the original study \cite{Park_etal_2023} is defined as follows: 

\begin{equation}
\textnormal{CD}_5(t) = \frac{1}{n_t}\sum_{i}^{n_t} -2f_{it}b_{it} + f_{it}
\label{eq1}
\end{equation}

\noindent where $f_{it}$ is a binary variable (1 or 0) for including the future $n_t$ papers within the $5$–year window that cite ($f=1$) the focal paper $i$ in year $t$ or not ($f=0$). Similarly, $b_{it}$ is a binary variable indicating whether the future paper cites a predecessor of the focal paper. Each term in the sum varies from –1 to +1 and each paper published has equal weight in the calculation of the average CD$_5(t)$ for year $t$.  Note how equation \ref{eq1} is an average across all papers in year $t$, such that a paper cited a few times counts as much towards the annual disruption score CD$_5$ as a paper cited thousands of times.  

As we assume instead that highly-cited papers contribute more to disruption, here we re-analyze the trend described in \cite{Park_etal_2023} using their dataset of millions of scientific publications. For each focal paper, we calculate disruption via a citation-weighted measure, as follows:

 \begin{equation}
    m\textnormal{CD}_5(t) = \frac{c_{it}}{n_t}\sum_{i}^{n} -2f_{it}b_{it} + f_{it}
\label{eq2}
\end{equation}

\noindent where $c_{it}$, the average number of citations of the focal paper per year, captures the magnitude of use of the focal paper \cite{Funk_etal_2017}. When we sum up the $m\textnormal{CD}_5(t)$ for all papers in year $t$ from equation (eq. \ref{eq2}), we then need to divide by the sum of citations across all papers in year $t$, a number which also increased exponentially over the eight decades (Fig.~\ref{fig1}, red).

In order to calculate this from the data available in the original study, \cite{Park_etal_2023} we use the CD$_5(t)$ metric and the average number of times $c_it$ each paper is cited each year: 

\begin{equation}
    m\textnormal{CD}_5(t) = \frac{1}{C_t}\sum_{i}^{N_t}c_{it}\textnormal{CD}_5(i)
\label{eq3}
\end{equation}

 \noindent where $C_t$ is the sum of all the citation counts $c_{it}$ of the $N_t$ papers published in year $t$. Normalizing in this way helps this weighted measure to account for the exponential growth in the total number of papers and citations over time. With this normalization, and because the disruption score in the sum ranges from –1 to +1, our null expectation is that equation (\ref{eq3}) should be close to zero in most years.  
 \begin{figure}[h!]
\centering
\includegraphics[width=.7\textwidth]{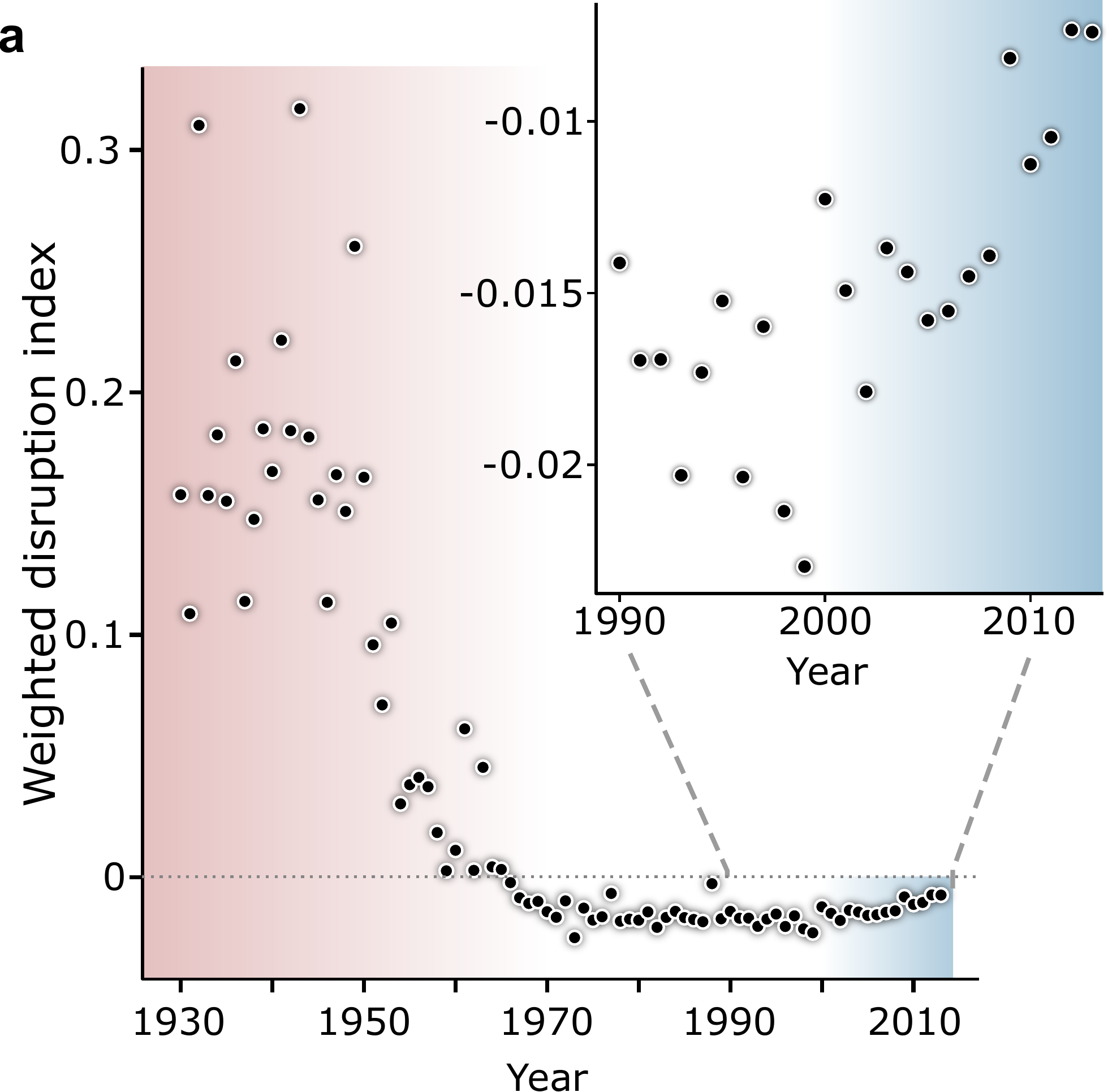} 
\caption{\textbf{The citation-weighted disruption index has been close to zero since around 1970}. Inset shows a modest increase in since 2000 (blue area). Source data from \cite{Park_etal_2023}.} 
    \label{fig2}
\end{figure}

This exponential increase in information obscures the overall pattern of disruption. By incorporating citation rate into the disruption score for each publication, we might better estimate the disruptive impact. Here we show how an impact-based measure of disruption, weighted by overall citation count, affects the overall picture. Measured in the dataset from the study \cite{Park_etal_2023}, the new citation-weighted measure (\ref{eq2}) has been close to zero for the fifty years after about 1970 (Figure \ref{fig2}). Zooming in, we see a modest but consistent growth in weighted disruption since 2000 (Figure \ref{fig2}, inset), which coincides closely with the digital age of publishing \cite{Evans_2008}, suggesting that technology is a main force behind disruption.

\begin{figure}[h!]
\centering
\includegraphics[width=.75\textwidth]{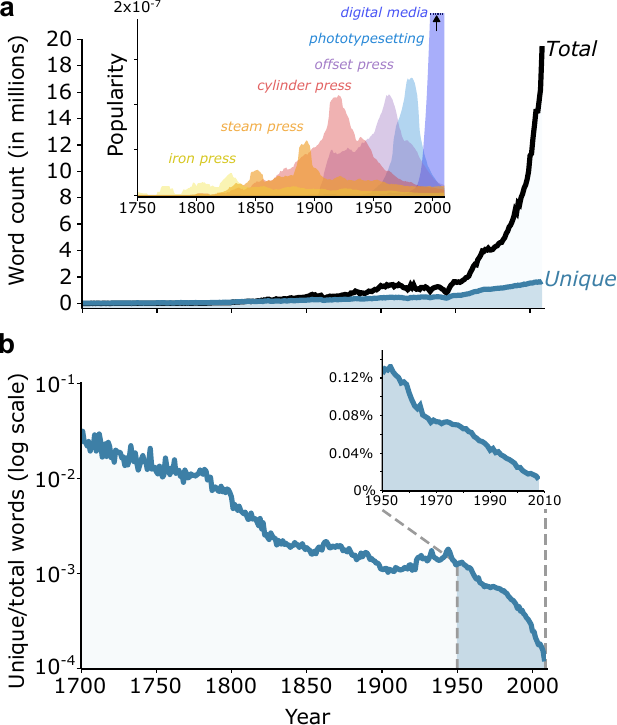} 
\caption{ \textbf{Corpus growth, fashion cycles and dilution.} \textbf{(a)} Annual count of words (black curve), in billions, and vocabulary (unique words, blue curve), in tens of millions, in English-language books. Inset: popularity cycles of selected technological terms (the max of "digital media" lies above the plot). \textbf{(b)} Unique words (appeared once in the year) as a fraction of total word count, with logarithmic y-axis (inset: 1950–2008, not logarithmic). All data are from the 2012 Google books database (https://booksgooglecom/ngrams/info).} 
    \label{fig3}
\end{figure}

The other effects observed in the study\textemdash the decrease in unique words, as well as changes in word frequency\textemdash appear to be mostly due to the accelerated growth in the number of publications. This is seen across modern information in general \cite{Hidalgo_2015, Duran-Nebreda_etal_2022}, as shown by the word count in English-language books over time \cite{Michel_etal_2010, Perc_2012} in Figure \ref{fig3}a.  Although we do not fully understand why things grow exponentially, many technological advances can accelerate cultural changes \cite{Bentley&OBrien_2017} and promote the need for novel words \cite{Duran-Nebreda_etal_2022} (Figure \ref{fig3}a, inset).  
In general, the ratio of unique terms to total words used declines as information grows (Figure \ref{fig3}b). This trend has also been reported in scientific publications and in books published in different languages across very different domains \cite{Petersen_etal_2012}. Similarly, the most-used verbs in science and patents have changed over time, reflecting common English language usage, and this pattern can be easily predicted by simple models of word choice \cite{Simon_1955, Ruck_etal_2017, Vidiella_etal_2022}. For example, the use of the word "produce" has declined over time both in patents and in English books in general (see Fig. 3 in \cite{Park_etal_2023}). 

In summary, we see little evidence that disruption is decreasing. The disruption metric is going down, but the citation-weighted measure is still going up. With each paper equally weighted regardless of impact, the measure of disruption with number of papers in the denominator (eq. \ref{eq1}) is expected to decrease over time amid an exponentially expanding corpus, as should the fraction of unique words. Instead, when we use the citation-weighted measure (eq. \ref{eq3}), we find that the disruption measure moves close to its expected value of zero over time, but also increases modestly in the last 20 years. Given this is amidst an exponentially growing corpus, we might even conclude that disruption is increasing in the current age of digital publishing. In the near future, machine learning and natural language processing tools may accelerate disruption even more \cite{Bentley_etal_2021}. 

\end{document}